\documentclass[osajnl,twocolumn]{revtex4}  %% REVTeX 4.0

\usepackage{graphicx}

\begin{document}

\title{Depolarization of backscattered linearly polarized light}

\author{L.F. Rojas-Ochoa$^{(1)}$, D. Lacoste$^{(2)}$, R.
Lenke$^{(3)}$, P. Schurtenberger$^{(1)}$ and F.
Scheffold\footnote{Corresponding author. E-mail:
Frank.Scheffold@unifr.ch}$^{(1)}$ }

\affiliation{$^{(1)}$ Department of Physics, University of
Fribourg, CH-1700 Fribourg, Switzerland,}

\affiliation{$^{(2)}$ Physico-Chimie Th\'eorique ESPCI, 10 Rue
Vauquelin, 75231 Paris Cedex 05, France}

\affiliation{$^{(3)}$ Fakult\"{a}t f\"{u}r Physik, Universit\"{a}t Konstanz,
D-78547 Konstanz,Germany\\ and Carl Zeiss Laser Optics GmbH,
D-73446 Oberkochen, Germany}

\date{\today}

\begin{abstract}
 We formulate a quantitative description of backscattered
linearly polarized light using an extended photon diffusion
formalism taking explicitly into account the scattering anisotropy
parameter $g$ of the medium. From diffusing wave spectroscopy
measurements the characteristic depolarization length for linearly
polarized light $\ell_p$ is deduced. We investigate the dependance
of this length on the scattering anisotropy parameter g spanning
an extended range from -1 (backscattering) to 1 (forward
scattering). Good agreement is found with Monte Carlo simulations
of multiply scattered light.

\end{abstract}


\maketitle

\section{Introduction}

Polarized light scattered many times in a random medium leaves the
sample partially depolarized.  Unfortunately, despite its
importance in areas like biomedical optical imaging, coherent
backscattering or dynamic spectroscopy
\cite{Maret87,DWS88,DWSReview,LenkeR}, the depolarization of light
in a random medium is still not completely understood due to the
complexity of vector wave multiple scattering (as compared to the
much simpler problem of scalar wave propagation). Previous
attempts have mainly focussed on isotropic (Rayleigh) scattering
\cite{Akkermans}, or on the depolarization of circularly polarized
light \cite{Gorodnichev}. Only few studies have discussed
specifically the mechanism of depolarization of linearly polarized
light in the case where the anisotropy parameter $g=<\cos \theta>$
is different from $0$ and furthermore a detailed comparison with
experiment has been lacking
\cite{MackJ,MackZ,Bicout,Lacoste2004,Zim01,Keller}. An accurate
description for arbitrary scattering anisotropy is however crucial
to analyze the information contained in backscattered light if
progress is to be made in applications like remote sensing, photon
correlation spectroscopy or optical imaging of biological tissues
\cite{DWSReview,Sebbah,Moscoso}.

 \begin{figure}[h]\centerline{\scalebox{1}{\includegraphics{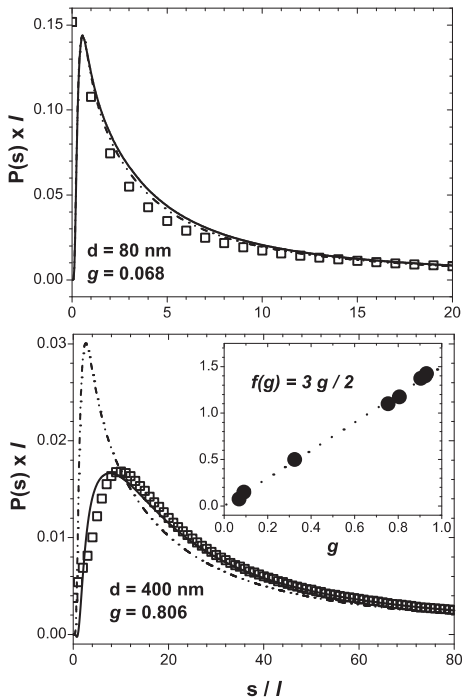}}}
  \caption{Normalized path length distribution $P(s) \times l$ for backscattered light from a semi-infinite medium. Symbols: Monte Carlo simulations. Lines:
    calculations based on Eq.~\ref{Pathd2} with $f(g)=3g/2$ (solid) and $f(g)\equiv 0 $
    (dashed). Inset: $f(g)$ obtained from Eq.~\ref{Pathdl} adjusted to fit the simulation
    results. Wavelength $\lambda_{0} = 532 \, nm$, refractive
    index of the particle $n_{p} = 1.59$ and of the solvent $n_{s} = 1.332$.
    Non-reflecting boundary conditions were used.}
  \label{fig1}
  \end{figure}

In this paper we formulate a quantitative description of
backscattered linearly polarized light using an extended photon
diffusion formalism taking explicitly into account the scattering
anisotropy parameter $g$. The details of our model are adjusted by
comparison with  Monte Carlo simulations of multiply scattered
light. We show how the characteristic length of depolarization of
incident linearly polarized light $\ell_p$ can be deduced from
measurements of intensity fluctuations of light scattered from
liquid turbid media via Diffusing Wave Spectroscopy (DWS). We can
distinguish the following limiting situations for the transport of
light and its polarization: isotropic scattering $g \simeq 0$,
forward-peaked scattering $g \simeq 1$ and backward-peaked
scattering $g \simeq -1$. The situation of forward-peaked
scattering $g \simeq 1$ is typical of Mie scattering \cite{Hulst}
with large particles and of biological tissues, whereas the
situation of $g<0$ has only been made possible experimentally
recently by tuning the interaction of the light using
mesostructured colloidal liquids \cite{LuisPRL2003}. Here we show
that with our additional correction the simple photon diffusion
picture successfully describes the distribution of path lengths
and the DWS autocorrelation function in the backscattering
geometry.

\section{Path lengths distribution for backscattering}

On length scales much larger than the transport mean free path
$l^*$, the transport of light in a turbid medium can be described
by the diffusion approximation. This approximation is connected to
the idea of treating the transport of photons as a random walk,
characterized by a distribution of path lengths
\cite{Akkermans,Pinerefle,RefleD,Ishimaru}. An exact solution of
the diffusion equation applied to light transport can be obtained
using the method of images. This method takes into account the
boundary conditions through two lengths which are both of the
order of a transport mean free path: the extrapolation length
$z_e$ where the flux of the flux of photons vanishes outside the
sample and $z_p$ which is the location of the photon source (for a
more detailed interpretation of $z_e,z_p$ based on a random walk
model see reference \onlinecite{LenkeR}). The method leads to
\begin{equation}\label{Pathdl}
    P(s) = \frac{\sqrt{3}}{4 \, \sqrt{\pi \, \ell^{\ast}} \, s^{3/2}}
    \left[ z_{p} \, e^{-\frac{3}{4} \, \frac{z^{2}_{p}}{\ell^{\ast} \, s}} +
    (z_{p} + 2 \, z_{e}) \, e^{-\frac{3}{4} \, \frac{(z_{p} +
    2 \, z_{e})^{2}}{\ell^{\ast} \, s}} \right].
\end{equation}
which obeys the normalization condition $\int_0^\infty P(s) ds=1$.
Note that the path length is simply related to the number of
scattering events $n$ by $s/\ell=n-1$, so that the path length is
$0$ for single scattering. Here $l$ is the scattering mean free
path. Both quantities, $l$  and  $l^*$, are related by $l^*/l =
1/(1 - g)$. The scattering anisotropy parameter $g$ is defined as
the average of the cosine of the scattering angle $g =
\left\langle {\cos \Theta } \right\rangle$.

To check the validity of Eq.~\ref{Pathdl}, we have performed
Monte-Carlo simulations of linearly polarized light reflected from
a semi-infinite turbid medium (details about the simulation method
can be found in refs. (\onlinecite{LenkeR} and
(\onlinecite{Lenke1}). These simulations use the Mie scattering
cross section in the range $0 \le g \le 1$ and are able to
evaluate numerically an exact path length distribution as a
function of the number of scattering events $n$ and polarization.
The simulations were done for uncorrelated spherical scatterers
(structure function $S(q) \equiv 1 $) and for a non-reflecting
interface characterized by $z_{p} \approx \ell^{\ast}$ and $z_{e}
\approx 0.7 \, \ell^{\ast}$. Values of $\lambda_{0}/n_s = 532 \,
nm$ for the incident wavelength, $n_{p} = 1.59$ for the refractive
index of the particle  and $n_{s} = 1.332$ for solvent refractive
index were used. In Fig. \ref{fig1} we compare the results of the
simulations with the prediction of the method of images according
to Eq.~\ref{Pathdl}. We see clearly in this figure that the method
of images provides an excellent description of the path length
distribution for the case of isotropic scattering ($g \equiv 0 $),
but that the method fails to give an equally good description for
the case of anisotropic scattering corresponding to $g \simeq
0.806$.

\begin{figure}[h]\centerline{\scalebox{1}{\includegraphics{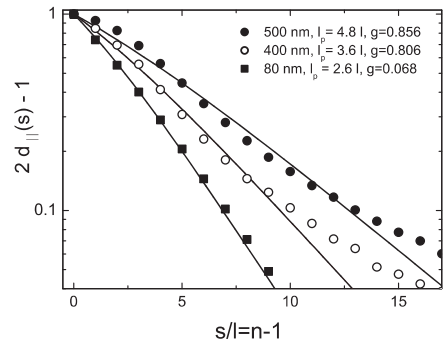}}}
  \caption{Depolarization of multiply scattered light. $2d_{\parallel}(s)-1$ from theory (Eq.(\ref{Psdep3}) solid lines)
  and simulation (symbols). Excellent agreement if found for Rayleigh scatterers
  while for larger particles the agreement becomes somewhat less good.}\label{fig2}
  \end{figure}

The disagreement in the latter case is not surprising as the
diffusion approximation is known to overestimate the contribution
from the short paths of the distribution, the error becoming more
and more severe as the anisotropy of scattering increases. One way
to improve the distribution of paths length of Eq.~\ref{Pathdl},
is by introducing a cutoff in the distribution as suggested by
Mackintosh and John \cite{MackJ}. Here we extend their approach by
taking into account explicitly the scattering anisotropy factor
$g$ (with $\int_0^\infty P_{corr}(s) ds=1$):
\begin{equation}
P_{corr}(s) \propto P(s) \cdot \left[ 1 - f(g) \, e^{- s /
\ell^{\ast}} \right] \ \label{Pathd2} \end{equation}

According to Mackintosh and John \cite{MackJ} $f(1)$ is of order
unity and for isotropic scattering ($g \rightarrow 0$) there is no
correction $f(0) = 0$. Using the corrected distribution of path
lengths to fit the simulation results, we have found that the
function $f$ is well approached by a linear dependance $f(g)
=3g/2$ which we assume to be valid also for $g<0$. As can be seen
in figure \ref{fig1}, the use of the correction factor
significantly improves the prediction of Eq.~\ref{Pathdl}. We note
that, alternatively, for the case of forward-peaked scattering
$g\simeq 1$, other schemes of approximations have been suggested.
For example the recent Ref.~\cite{Keller} reports that the
Fokker-Plank equation provides a better description than does the
(uncorrected) diffusion approximation for forward-peaked
scattering.

\goodbreak

\section{Depolarization length for linear polarization}\label{Depo}

Incident polarized light looses its polarization in random
multiple scattering \cite{Akkermans,MackJ,MackZ,Bicout,Zim01}. For
linearly polarized light, only two configurations ($\parallel$ and
$\perp$) need to be considered (for isotropic samples).
Physically, in the $\parallel$ geometry more photons are detected
for short paths as compared to the unpolarized case. In a seminal
paper Akkermanns et al. \cite{Akkermans} found that the path
length distribution for the two configurations can be written as
\begin{equation}
  P_{\parallel,\perp}(s)= d_{\parallel,\perp}(s) \cdot P(s)\label{akkdep}
\end{equation}
with the depolarization ratio given by
\begin{equation}\label{Psdep3}
d_{\parallel}(s)=\frac{{1 + 2e^{ - s/\ell_p } }}{{2 + e^{ -
s/\ell_p }, }}
\end{equation}
\begin{equation}\label{Psdep4}
d_{\perp}(s)=\frac{{1 - e^{ - s/\ell_p } }}{{2 + e^{ - s/\ell_p }
}},
\end{equation}
in terms of the characteristic length of depolarization for
linearly polarized light $\ell_p$. For point like scatterers ($g =
0 $) Akkermans et al. obtained $\ell_{p}= \ell / \ln (10 / 7)
\cong 2.804 \, \ell$ \cite{Akkermans}. We find good agreement
between Eq.\ref{akkdep} and our numerical simulations with
$\ell_p$ as an adjustable parameter (Fig.\ref{Fig2}). For large
particles (and therefore large $g$) the agreement is somewhat less
good. However polarization effects in DWS usually are found weak
for $g \approx 1$ and therefore we did not attempt to improve the
accuracy of Eqs. (\ref{Psdep3}) and (\ref{Psdep4})(It is
worthwhile to note that  close to the sample surface very
interesting polarization effects, such as butterfly pattern,
persist \cite{Writhing,Hielscher97}).

In the limit $s/\ell \gg 1$ Eq. \ref{Psdep3} and \ref{Psdep4}
reduce to :
\begin{equation}\label{Psdep5}
P_{\parallel,\perp}(s) \cong  \left[ \frac{1}{2} \pm \frac{3}{4}
\, e^{-s/\ell_{p}}
  \right] \cdot P(s)
\end{equation}

We consider this  expression the most simple generalization since
it captures well intermediate path lengths $s/\ell > 3$, where
polarization effects are important, but at the same time the
number of scattering events is already sufficiently large to apply
the diffusion approximation.

%%% ----------------------------------------------------------------------

\section{Polarization dependence of DWS autocorrelation function}
In transmission geometry the path length distribution can be
measured experimentally using pulsed laser beams\cite{yodh90}. In
reflection however paths are short and therefore the time
resolution is usually not sufficient for such measurements. An
alternative way to probe diffuse light propagation is the analysis
of temporal fluctuations of the scattered light via photon
correlation spectroscopy. This approach, called diffusing wave
spectroscopy (DWS), is a sensitive probe to the path length
distribution, in particular in reflection
geometry\cite{Maret87,DWS88}. The temporal intensity correlation
function $g_2 (t)$ which is measured, is related to the field
autocorrelation function by the Siegert relation
$g_1(t)=\sqrt{1-g_2(t)}$. The latter is directly related to the
path length distribution:
\begin{equation}
g_1 (t) = \int\limits_0^\infty  {P(s)e^{ - 2\left( {t/\tau_0 }
\right)s/\ell^{\ast }} ds} \label{g1full}.
\end{equation}
In our case the characteristic relaxation time for diffusive
particle motion (diffusion constant $D$ for free Brownian motion)
$\tau _0 = \left( {k_0^2 D} \right)^{ - 1}$ is a known quantity.
The path length distribution P(s), as given by Eq.~(\ref{Pathdl}),
has been derived for the case of a semi-infinite non-absorbing
medium. For real systems, both absorption and limited container
size lead to a loss of photons along a given path. For such cases
the distribution of path length becomes $P'(s) = \exp (-s/\ell_a
)P(s)$, where $\ell_a$ is the characteristic absorption length of
the medium. In this framework, absorption can be taken into
account by the modification

\begin{equation} 6t/\tau _0 \to 6t/\tau_0 + 3 \ell^*/\ell_a
\label{absorbtioncorrection}\end{equation}

in Eq.~\ref{g1full} \cite{DWSReview}. All our experiments
correspond to a case where $\ell_a$ is much larger than
$\ell^{\ast}$.

The solution for the scalar (polarization independent) path length
distribution is well known \cite{DWSReview} (neglecting
absorption): $ g_{1}(t) = \left[ e^{- \gamma_{p} \, x(t)} + e^{-
(\gamma_{p} + 2 \, \gamma_{e}) \, x(t)} \right]/2$, where $x(t) =
\sqrt{6 \, t / \tau_{0}}$, $\gamma_{p} = z_{p} / \ell^{\ast}$ and
$\gamma_{e} = z_{e} / \ell^{\ast} \, $. In the limit $x \ll 1$ it
reduces to
\begin{equation}\label{g1scalar2}
g_{1}(t) = e^{- \gamma \, x}
\end{equation}
with $\gamma = - \frac{\partial \, \ln g_1}{\partial \, x} (x=0) =
\gamma_{p} + \gamma_{e}$. Note that this expression is independent
of $\ell^*$ and that for the non-reflecting interface that we have
considered $\gamma = 1 + z_e/\ell^*=5/3$. All different paths of
length $s$ contribute to $g_1(t)$. Short paths predominantly
contribute to the long time decay and long paths contribute to the
short time decay. Clearly polarization effects modify the path
length distribution and therefore strongly influence the decay of
$g_1(t)$. Interestingly the shape of $g_1(t)$ remains more or less
unchanged. In most previous studies Eq.~\ref{g1scalar2}, though
derived for the scalar case, has been applied also with polarized
light. Although $\gamma$ is in principle a well defined constant,
it has been treated in the literature as an adjustable parameter
to explain the polarization dependence of the correlation
function. Values of $\gamma_{\perp,\parallel}$ in a range $1$ to
$3$ have been reported depending on detected polarization state,
particle size and concentration
\cite{DWSReview,MackZ,LuisPRE2002}.

Rather than to adjust the parameter $\gamma$ we take into account
the polarization via Eq.~\ref{Psdep5} and find:

\begin{eqnarray}\label{cfdp}
    \nonumber
    g_{1 \, \perp}(t) &=& \frac{\int \, P_{\perp}(s) \, e^{- 2 \, (t / \tau_{0})
    \, s / \ell^{\ast} \, ds}}{P_{\perp}(s) \, ds} \\
    & & \\
    \nonumber
    g_{1 \, \parallel}(t) &=& \frac{\int \, P_{\parallel}(s) \, e^{- 2 \, (t / \tau_{0})
    \, s / \ell^{\ast} \, ds}}{P_{\parallel}(s) \, ds} \, .
\end{eqnarray}\\
Introducing the function $h(x)=(e^{(-\gamma_p x)}+e^{
-(\gamma_p+2\gamma_e )x})/2$, the correlation functions take the
form (neglecting absorption)
\begin{equation}\label{depcfc}
g_{1,\perp}(t)=\frac{h \left[x_1(t) \right]- \frac{3}{2} h \left[
y_1(t) \right] -\frac{3g}{2} h \left[x_2(t) \right]+ \frac{9g}{4}
h \left[ y_2(t) \right] } {h \left[x_1(0) \right]- \frac{3}{2} h
\left[ y_1(0) \right] -\frac{3g}{2} h \left[x_2(0) \right]+
\frac{9g}{4} h \left[ y_2(0) \right] },
\end{equation}
\begin{equation}\label{depcfc2}
g_{1,\parallel}(t)=\frac{h \left[x_1(t) \right]+ \frac{3}{2} h
\left[ y_1(t) \right] -\frac{3g}{2} h \left[x_2(t) \right]-
\frac{9g}{4} h \left[ y_2(t) \right] } {h \left[x_1(0) \right]+
\frac{3}{2} h \left[ y_1(0) \right] -\frac{3g}{2} h \left[x_2(0)
\right]- \frac{9g}{4} h \left[ y_2(0) \right] },
\end{equation}
where $x_{1}(t) = \sqrt{6 \, t / \tau_{0}}$, $x_{2}(t) = \sqrt{6
\, t / \tau_{0} + 3}$, $y_{1}(t) = \sqrt{6 \, t / \tau_{0} + 3 \,
\ell^{\ast} / \ell_p}$ and $y_{2}(t) = \sqrt{6 \, t / \tau_{0} + 3
+ 3 \, \ell^{\ast} / \ell_p} \, $. This set of equations provide a
direct relation between measurements detecting $\perp$ or
$\parallel$ polarized light thereby eliminating the need to
introduce two adjustable parameters $\gamma_{\perp, \, \parallel}
\,$. We carried out a series of dynamic multiple scattering
experiments to follow the polarization memory of the reflected
light intensity. Experiments were realized as described in
\cite{LuisPRE2002} but with $\lambda=532$nm. The sample cells were
suspended in a water bath to suppress reflections and to maintain
a constant temperature of $T=22^\circ C$. All samples used were
made from monodisperse polystyrene particles (n=1.595) suspended
in water (n=1.332), except for one case ($d=114nm$) where we used
a mixture of water
and ethanol \cite{LuisPRL2003} (n=1.365). A detailed description of all samples is given in table \ref{Sizesgs} \\
\[
\;
\]

\begin{table}[t]
    \centering
    \begin{tabular}{|r|c|c|}
    \hline
    d  ($nm$)   & $\Phi$ ($\%$)& $g$\\

    \hline
      80 $\pm$ 5 & 3.9                 & 0.04 \\
      92 $\pm$ 15 & 4.1                 & 0.05 \\
     168 $\pm$ 5 & 2.0 $\to$ 30.0         & 0.29 $\to$ -0.13 \\
     350 $\pm$ 12 & 1.9                & 0.74 \\
     400 $\pm$ 8 & 1.9                & 0.80 \\
     720 $\pm$ 14 & 1.9               & 0.90 \\
    1000 $\pm$ 51 & 2.0               & 0.92 \\
    1500 $\pm$ 53 & 2.0               & 0.92 \\
     114 $\pm$ 10 & 2.5 $\to$  7.4        & -0.25 $\to$  -0.78 \\
    \hline
    \end{tabular}
    \caption{Polystyrene spheres used in the experiments. Diameter $d$ as
    obtained from dynamic light scattering, volume fraction $\Phi$ and  scattering anisotropy parameter $g$ . The latter was obtained either from Mie calculations (for hard sphere interactions as described in \cite{LuisPRE2002,LuisPRL2003}) or direct measurements (for the $d=114nm$ charged spheres as described in \cite{LuisPRL2003}) .}\label{Sizesgs}
\end{table}
\goodbreak

\begin{figure}[h]\centerline{\scalebox{1}{\includegraphics{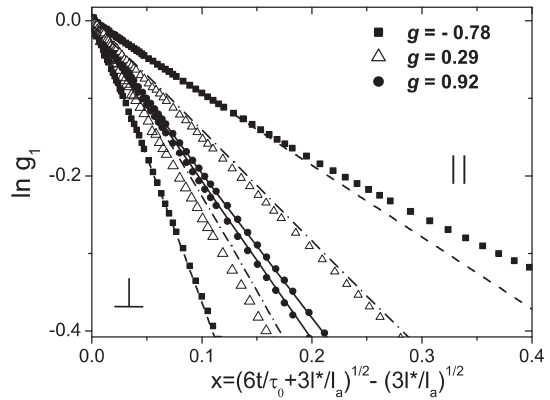}}}
  \caption{DWS autocorrelation
    function for different $g$ -values detecting polarized and depolarized
    light. Lines: calculations based on Eq. \ref{depcfc} with $\ell_p$ adjusted to best
fit the data. ($g=-0.78: \ell^*/\ell_a  = 0.004,
\ell_p/\ell^*=5.4$), ($g=0.32: \ell^*/\ell_a=0.0064,
\ell_p/\ell^*=2.21$), ($g=0.924, \ell^*/\ell_a=0.016,
\ell_p/\ell^*=0.713$). }\label{fig3}
  \end{figure}

In Fig. \ref{fig3} we show a comparison between experiments and
our theoretical expressions. For clarity the data is normalized
and plotted as a function of $x=\sqrt{6t/t_0+3l^*/l_a}-
\sqrt{3l^*/l_a}$ thus removing contributions from absorption at
$x<0.1$ (Eq.(\ref{absorbtioncorrection})) (The absorption length
$l_a$ has been chosen such that $ln{g_1(x)}$ scales linearly at
small $x$-values). We find that the theory describes our data very
well, with the polarization length $\ell_p$ being the only
adjustable parameter. In particular the availability of
suspensions with negative $g$ values allows a rigorous test of the
model over the whole interval $-1 \le g \le 1$. Note that we use
the theoretically predicted values $\gamma_p,\gamma_e$ and do not
adjust them to fit the data (as done in all previous work).
Depolarization lengths obtained from fits (Eq.~\ref{depcfc}) to
the DWS data and from numerical simulations (Fig.\ref{Fig2} ) are
presented in Fig. \ref{fig4}. We note again that both DWS
experiments ($g_{1 \, \perp},g_{1 \,
\parallel}$) are well  characterized by a single $\ell_p$
even for the most extreme case of $g \to -1$.

In the limit $g \simeq 0$ we find good agreement with the
predicted theoretical value of Akkermans $\ell_{p} / \ell \cong
2.804$ \cite{Akkermans}. As the anisotropy parameter $g$ is
increased $\ell_{p} / \ell$ slowly increases as well. In the case
of forward-peaked scattering $g\simeq 1$, we find $\ell_p \simeq
\ell^*$ (see also refs. (\onlinecite{Bicout}) and
(\onlinecite{Lacoste2004})) . This means that as $g$ approaches
$1$ an increasing number of scattering events is necessary to
depolarize backscattered light. Since $\ell_p/\ell^*$ remains
constant for $g \to 1$ the ratio $\ell_p/\ell$ has to increase
sharply, as shown in Figure \ref{fig4}b. In the case of
backward-peaked scattering $g \simeq -1$ however, the number of
scattering events needed to depolarize remains virtually
unchanged. The characteristic length scale of depolarization is
still the scattering mean free path $\ell$ (and not $\ell^*$ !) as
in the case of point-like scatterers. Therefore it is not
surprising to see that in Fig. \ref{fig4} $\ell_p$ scales as in
the Rayleigh limit {\it i.e} $\ell_p \simeq 3 \ell$ even for a
transport mean free path $\ell^* \simeq \ell/2$. When
$\ell_p/\ell^*$ is plotted as a function of $g$ we find an almost
linear behavior over the full accessible range. At this point our
understanding of $\ell_p$ is limited to the particular cases
($g=0$, $g=1$, and $g=-1$), a more detailed microscopic approach
would be required to explain the complete dependance of $\ell_p$
on $g$, or the dependance on other single particle optical
properties.

\section{Depolarization ratio of backscattered intensities}

An alternative way to study polarization of multiple backscattered
waves is through the (full intensity) depolarization ratio:

\begin{equation}\label{drat1}
    d = \frac{I_{\parallel} - I_{\perp}}{I_{\parallel} +
    I_{\perp}}\, .
\end{equation}

\begin{equation}\label{drat2}
    d = \frac{\int \, [P_{\parallel}(s) - P_{\perp}(s)] \, ds}
    {\int \, [P_{\parallel}(s) + P_{\perp}(s)] \, ds} \cong
    \int_{0}^{\infty} \frac{3}{2} \, \, e^{- s / \ell_p} \, P(s) \, ds \, ,
\end{equation}
from where we identify a $s$-dependent depolarization ratio
$d(s)$:

\begin{equation}\label{ds}
    d(s) \cong \frac{3}{2} \, e^{- s / \ell_p} \, .
\end{equation}
Since $P_{\parallel}(s) + P_{\perp}(s) = P(s)$ and
$\int_{0}^{\infty} \, P(s) \, ds = 1$, therefore:

\begin{eqnarray}\label{drat3}
    \nonumber
    d &=& \int_{0}^{\infty} \, P(s) d(s) \, ds \\
     & & \\
    \nonumber
     &\cong& \frac{3}{4} \, \left( e^{- \gamma_p \, \sqrt{3 \, \ell^{\ast} / \ell_p}} +
    e^{- (\gamma_{p} + 2 \, \gamma_{e}) \, \sqrt{3 \, \ell^{\ast} / \ell_p}} \right) \, .
\end{eqnarray}

\begin{figure}[h]\centerline{\scalebox{1}{\includegraphics{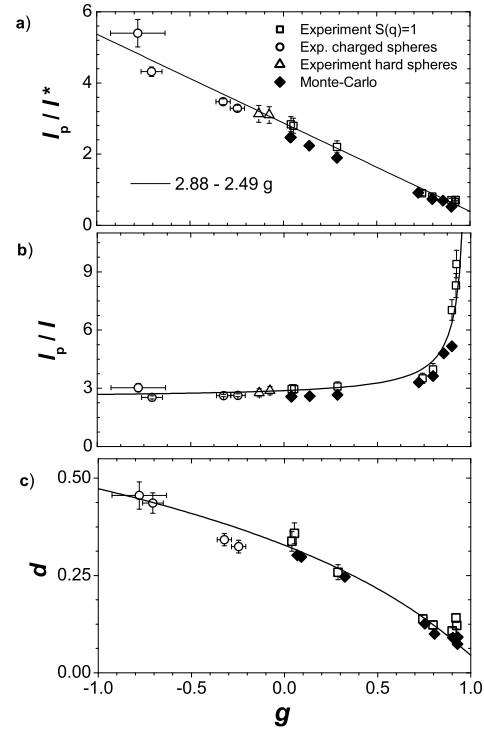}}}
  \caption{a) and b) Depolarization
    length $\ell_p$ from DWS measurements and Monte-Carlo simulations. Squares: measurements for different
    particle sizes (random particle configuration $S(q) \equiv  1$).
    Circles: Strongly interacting charged particles \cite{LuisPRL2003}. Triangles:
   Hard sphere data from reference \onlinecite{LuisPRE2002}. Diamonds: Monte-Carlo simulations.
   c) Depolarization ratio directly obtained
    from the measured intensities. Solid lines: calculated from the linear fit
    to $\ell_p/\ell^*$ shown in panel a) \cite{Lenke1,LenkeR}.}
\label{fig4}
  \end{figure}

\goodbreak
\noindent In Figure \ref{fig4}c we show the experimental values of
$d$. Again the agreement is excellent over the full accessible
range. The data comes very close to the predicted values in the
three particular cases: $d=0.33$ for Rayleigh scattering ($g=0$),
$d=0.14$ for forward-peaked scattering ($g=1$) and $d=0.49$ for
backward peaked scattering ($g=-1$).

\section{Conclusion}

In this paper, we have shown how to describe the effect of the
scattering anisotropy on the depolarization of linearly polarized
light, and how to use Diffusing Wave Spectroscopy (DWS) to
determine the characteristic depolarization properties. By means
of numerical simulations, we checked the limit of validity of the
diffusion approximation when the scattering anisotropy $g$ is
increased, and we have shown how to correct the predictions by
means of an anisotropy dependent cutoff for the path length
distribution $P(s)$. We discuss for the first the time the
dependence of the characteristic depolarization length over the
full range of possible values of $g$ including the unusual case of
negative $g$ values. In our description the extrapolation length
$\gamma$ is a well defined constant as required by diffusion
theory. Our work thus clarifies the meaning of $\gamma$, subject
to intense discussion in the past \cite{Freund,Rosenbluh}. Since
our description only uses a single adjustable parameter, it is now
possible to fully characterize backscattered light with
polarization resolved measurements. We think that this approach
can strongly benefit applications in the field of soft and bio
material analysis, as well as diffuse light imaging techniques
\cite{DWSReview,Scheffold2002,Keller}.

%%% ----------------------------------------------------------------------
\goodbreak

\acknowledgments We thank D. Pine, M. Cloitre and A. Maggs and F.
Jaillon for discussions. Financial support from the Swiss National
Science foundation is gratefully acknowledged.

\newpage


\begin{thebibliography}{99}


\bibitem{DWSReview} D. A. Weitz and D. J. Pine, in \emph{Dynamic Light Scattering},
ed. by W. Brown, Oxford U. Press; New York(1993), Chap. 16, pp.
652-720.

\bibitem{Maret87} G. Maret and P.E. Wolf, "Multiple Light-Scattering From Disordered Media - The Effect Of Brownian-Motion Of
Scatterers", Z. Phys. B {\bf 65}, 409-431 (1987).

\bibitem{DWS88} D. J. Pine, D. A. Weitz, P. M. Chaikin, and E. Herbolzheimer,
"Diffusing-Wave Spectroscopy", Phys. Rev. Lett. {\bf 60},
1134-1137 (1988).

\bibitem{LenkeR} R. Lenke and G. Maret, in \emph{Multiple Scattering of Light:
Coherent Backscattering and Transmission}, ed. by W. Brown, Gordon
and Breach Science Publishers; Reading U. K. (2000), pp. 1-72.

\bibitem{Akkermans} E. Akkermans, P. E. Wolf, R. Maynard and G. Maret,
"Theoretical-Study Of The Coherent Backscattering Of Light By
Disordered Media", J. Phys. France {\bf 49}, 77-98 (1988)

\bibitem{MackZ} F. C. MacKintosh, J. X. Zhu, D. J. Pine and D. A. Weitz,
"Polarization Memory Of Multiply Scattered-Light",
 Phys. Rev. B {\bf 40}, 9342-9345 (1989).

\bibitem{MackJ} F. C. MacKintosh and S. John,
"Diffusing-Wave Spectroscopy And Multiple-Scattering Of Light In
Correlated Random-Media", Phys. Rev. B {\bf 40}, 2383-2406 (1989).

\bibitem{Bicout} D.~Bicout, C.~Brosseau, A.~S. Martinez, and J.~M.
Schmitt, "Depolarization of multiply scattered waves by spherical
diffusers: Influence of the size parameter", Phys. Rev. E, {\bf
49}, 1767-1770, (1994).


\bibitem{Lacoste2004}D. Lacoste, V. Rossetto, F. Jaillon and H.
Saint-Jalmes, submitted

\bibitem{Zim01} D. A. Zimnyakov, Y. P. Sinichkin, P. V. Zakharov and D. N. Agafonov,
"Residual polarization of non-coherently backscattered linearly
polarized light: the influence of the anisotropy parameter of the
scattering medium", Waves Random Media {\bf 11}, 395-412 (2001).

\bibitem{Keller}
A.~D. Kim and J.~B. Keller, "Light propagation in biological
tissue", J. Opt. Soc. Am. A, {\bf 20} (1), 92-98, (2003).


\bibitem{Gorodnichev}
E. E. Gorodnichev, A. I. Kuzovlev and D. B. Rogozkin, "Diffusion
of circularly polarized light in a disordered medium with large-
scale inhomogeneities", JETP Letters, {\bf 68}(1), 22-28, (1998).

\bibitem{Moscoso}
M. Moscoso, J. B. Keller and G. Papanicolaou, "Depolarization and
blurring of optical images by biological tissue", J. Opt. Soc. Am.
A, {\bf 18} (4), 948-960, (2001).

\bibitem{Sebbah} P. Sebbah, Waves and imaging through complex media (Kluwer
Academic Publishers, Dordrecht ; Boston, 2001).

\bibitem{Hulst} H. C. van de Hulst, \emph{Light Scattering by Small
Particles}, Dover, New York (1981). C. F. Bohren and D. R.
Huffman, \emph{Absorption and Scattering of Light by Small
Particles}, Wiley, New York (1983).

\bibitem{LuisPRL2003}  L.F. Rojas-Ochoa, J.M. Mendez-Alcaraz, J.J. Saenz,
P. Schurtenberger and F. Scheffold, "Photonic properties of
strongly correlated colloidal liquids", submitted; We note that in
present work we could not access concentrations above $\Phi = 7.4
\%$ due to the apparent non-egodicity of the scattering signal
being much more pronounced in backscattering as compared to the
transmission geometry studied before.


\bibitem{Ishimaru} A. Ishimaru, \emph{Wave propagation and scattering in
random media}, Academic Press, New York (1978).

\bibitem{RefleD} A. Lagendijk and R. Vreeker and P. DeVries, "Influence Of
Internal-Reflection On Diffusive Transport In Strongly Scattering
Media", Phys. Lett. A {\bf 136}, 81-88 (1989).

\bibitem{Pinerefle} J. X. Zhu and D. J. Pine and D. Weitz, "Internal-Reflection
Of Diffusive Light In Random-Media", Phys. Rev. A {\bf 44},
3948-3959 (1991).

\bibitem{Lenke1} R. Lenke and G. Maret, "Magnetic field effects on coherent backscattering of light",
Eur. Phys. J. B {\bf 17}, 171-185 (2000).


\bibitem{Writhing} A. C. Maggs and V. Rossetto, "Writhing Photons and Berry
Phases in Polarized Multiple Scattering", Phys. Rev. Lett. {\bf
87}, 253901 (2001).


\bibitem{Hielscher97} A.H. Hielscher et. al., Optics exp. {\bf
1}, 441-453 (1997).

\bibitem{yodh90} A.G. Yodh, P.D. Kaplan, D.J. Pine, "Pulsed Diffusing-Wave Spectroscopy -
High-Resolution Through Nonlinear Optical Gating", Phys. Rev. B
{\bf 42}, 4744-4447 (1990).

\bibitem{LuisPRE2002} L. F. Rojas-Ochoa and S. Romer and F. Scheffold and P. Schurtenberger,
"Diffusing wave spectroscopy and small-angle neutron scattering
from concentrated colloidal suspensions", Phys. Rev. E {\bf 65},
051403 (2002).


\bibitem{Rosenbluh} M. Rosenbluh and M. Hoshen and I. Freund and M. Kaveh,
"Time evolution of universal optical fluctuations", Phys. Rev.
Lett {\bf 58}, 2754-2757 (1987).

\bibitem{Freund} I. Freund and M. Kaveh, "Comment on 'Polarization memory of multiply scattered light'",
Phys. Rev. B {\bf 45}, 8162-8163 (1992); F. C. MacKintosh and J.
X. Zhu and D. J. Pine and D. A. Weitz, "Reply to comment on
'Polarization Memory Of Multiply Scattered-Light'", Phys. Rev B
{\bf 45}, 8165-8165 (1992).

\bibitem{Scheffold2002} F. Scheffold, "Particle sizing with diffusing wave,
spectroscopy", J. Disp. Sci. Tech {\bf 23}, 591-599 (2002).




\end{thebibliography}
\end{document}